\documentclass{aastex62}

\usepackage{soul}

\usepackage{color}
\newcommand{\al}\textbf{}

\usepackage{multirow}

\usepackage{amsmath}	
\usepackage{amssymb}	
\usepackage{gensymb}
\usepackage{subfigure}

\begin{document}

\title{Candidate Hypervelocity Red Clump Stars in the Galactic Bulge Found Using the VVV and Gaia Surveys \footnote{Based  on  observations  taken  within  the  ESO  programmes
  179.B-2002, and 198.B-2004.}}

\correspondingauthor{Alonso Luna}
\email{aluna@astro.unam.mx; alonso.luna.rf@gmail.com}

\author[0000-0001-5971-8058]{Alonso Luna}
\affiliation{Instituto de Astronom\'ia, Universidad Nacional Aut\'onoma de M\'exico\\
Ciudad Universitaria\\
04510, M\'exico.}
\affil{Depto.  de   Ciencias  F\'isicas,  Facultad   de  Ciencias  Exactas, Universidad  Andres Bello\\
Av. Fernandez  Concha 700\\
Las  Condes, Santiago, Chile.}

\author[0000-0002-7064-099X]{Dante Minniti}
\email{vvvdante@gmail.com}
\affil{Depto.  de   Ciencias  F\'isicas,  Facultad   de  Ciencias  Exactas, Universidad  Andres Bello\\
Av. Fernandez  Concha 700\\
Las  Condes, Santiago, Chile.}
\affiliation{Millennium  Institute  of  Astrophysics\\
Av.  Vicuna  Mackenna 4860\\
782-0436, Santiago, Chile.}
\affiliation{Vatican  Observatory\\
V00120 Vatican  City  State,  Italy.}

\author[0000-0003-3496-3772]{Javier Alonso-Garc\'ia}
\email{javier.alonso@uantof.cl}
\affiliation{Centro de Astronom\'ia (CITEVA), Universidad de Antofagasta\\
Av. Angamos 601\\
Antofagasta, Chile.}
\affiliation{Millennium  Institute  of  Astrophysics\\
Av.  Vicuna  Mackenna 4860\\
782-0436, Santiago, Chile.}

\begin{abstract}
We propose a new way to search for hypervelocity stars in the Galactic bulge, by using red clump (RC) giants, that are good distance indicators.  
The 2nd Gaia Data Release and the near-IR data from the VISTA Variables in the Via Lactea (VVV) Survey led to the selection of a volume limited sample of 34 bulge RC stars.
A search in this combined data set leads to the
discovery of seven candidate hypervelocity red clump stars in the Milky Way bulge.
Based on this search we estimate the total production rate of hypervelocity RC stars from the central supermassive black hole (SMBH) to be $N_{HVRC} = 3.26 \times 10^{-4} $ yr$^{-1}$.
This opens up the possibility of finding larger samples of hypervelocity stars in the Galactic bulge using future surveys, closer to their main production site, if they are originated by interactions of binaries with the central SMBH.

\end{abstract}


\section{Introduction} 
\label{sec:intro}

Hypervelocity stars (HVS) are rare objects that appear to be unbound from the Galaxy. For an authoritative and broad discussion about these objects see \citet{brown15}.
We concentrate on the HVS coming from the Galactic centre, produced by the interaction of the central SMBH with a binary system \citep{hills}. The Galactic bulge in particular is the region where one expects to find the highest density of hypervelocity stars produced by the suspected to be SMBH at the centre of our Galaxy.

Since the first discovery by \citet{brown05}, quite a few HVS have been  found  in the Milky Way halo with typical velocities $\sim 1000$ km s$^{-1}$ \citep[e.g.][]{brown08,brown14,brown15,brown18,Heber,kollmeier09,kenyon,palladino,Geier}, and also in  the Magellanic Clouds \citep[e.g.][]{Przybilla,lennon}. 

In order to explain the ejection process, different model predictions have been made besides the Hills one; \citet{yu} proposed a binary massive BH and a single star ejection mechanism; a globular cluster and a SMBH scenario was introduced by \citet{Fragione17} and a few others have been introduced by other authors \citep[e.g.][]{baumgardt,silk12,zhang,kenyon,rossi,FragioneLoeb17,marchetti,boubert,irrgang}.

On the basis of these previous studies, it has been estimated that every $\sim 10000$ yr one star is ejected by  binary interaction with the Galactic nuclear BH (Sag A*) as a HVS. If we can measure time of ejection, then we can check for bursts due to the accretion of a star cluster by the  central SMBH for example \citep{Capuzzo,Fragione16}.

In this paper we propose a new way to search for HVS in the Galactic bulge. We use the Gaia DR2 sample combined with the VVV sample to select bulge RC stars and search for HVS, reporting the discovery of two candidate hypervelocity RC stars in the Galactic bulge. 

\section{The Gaia and VVV Data}
\label{sec:sec2}

Because of the large non-uniform reddening in the bulge fields studied here \citep[e.g.][]{schlafly,gonzalez12,Minniti16,Minniti18,alonso17,alonso18}, we use the combination of optical Gaia data with near-IR VVV data.

The Gaia second data release (DR2) contains data collected between July 2014 and May 2016  \citep{gaia18}. It improves the photometric and astrometric measurements of the first release (DR1) as well as information on astrophysical parameters, variability and median radial velocities for some sources. Gaia DR2 contains the apparent brightness in $G$ magnitude for more than $1.6\times 10^9$ sources brighter than 21 mag and, for $1.4\times 10^9$ sources, broad band colours $G_{BP}$ covering $(330-680~nm)$ and $G_{RP}$ covering $(630-1050~nm)$, which were not available in DR1. The proper motion (PM) components in equatorial coordinates are available for $1.3\times 10^9$ sources with an accuracy of $0.06$ mas yr$^{-1}$ for sources brighter than $G=15$ mag, $0.2$ mas yr$^{-1}$ for sources with $G=17$ mag and $1.2$ mas yr$^{-1}$ for sources with $G=20$ mag \citep{gaia18}. The data have been processed by the Gaia Data Processing and Analysis Consortium (DPAC).


The VVV survey maps the Galactic Bulge and southern disk in the near-IR with the VIRCAM (VISTA InfraRed CAMera) at the 4.1m wide-field Visible and Infrared Survey Telescope for Astronomy \citep[VISTA,][]{emerson10} at ESO Paranal Observatory \citep{Minniti10,Saito12}. In the Galactic bulge, the VVV Survey covered 300 sq.deg. (within $-10\degree<l<10\degree$, $-10\degree<b<5\degree$), using the near-IR passbands: $Z (0.87~ \mu m),~ Y (1.02~ \mu m),~ J (1.25 ~\mu m), ~H (1.64~ \mu m)$, and $K_S (2.14 ~\mu m)$.

The VVV Survey data reduction and the archival merging were carried out at the Cambridge Astronomical Survey Unit \citep[CASU,][]{irwin}  and VISTA Science Archive (VSA) at the Wide-Field Astronomy Unit (WFAU), within the VISTA Data Flow System \citep{cross12}. 

In order to deal with the crowding in the VVV Survey field of view, we use the dataset of \citet{alonso18}, who built a new and more complete VVV photometric catalog using PSF photometry, which increased up to almost a billion the number of detected sources.

\section{The Selection of Hypervelocity RC stars}
\label{sec:sec3}

In order to select possible HVS candidates from the RC of the Bulge, we need to first define the appropriate region of the CMD in different appropriate colors where the RC lie. For that, and relying on the accuracy of the coordinates from both databases, we match the stellar position of the Gaia and VVV sources with a conservative tolerance of 0.5 arcsec (1.5 VVV pixels), resulting in a total of $N_{total}=29,181,380$ sources that have accurate positions in the Galactic bulge region (within $-10\degree<l<10\degree$, $-10\degree<b<5\degree$), optical and near-IR photometry and PMs.

The Wesenheit magnitudes were originally created for Cepheids \citep{Madore82}, but they can be applied to any kind of star; we adopted them in our study since they are very succesful in removing the reddening effect. The relations for the $K_S$ and $G$ band are respectively equations \ref{eq:wesenheit k} and \ref{eq:wesenheit g}: 

\begin{equation}
W_{K_S}= K_S - 0.45 \times (J-K_S), 
\label{eq:wesenheit k}
\end{equation}

\begin{equation}
W_G= G - 1.90 \times (G_{BP}-G_{RP})   . 
\label{eq:wesenheit g}
\end{equation}
\\

\begin{figure}[h]
\centering
\subfigure[]{%
\label{fig:cmd g-k}%
\includegraphics[width=0.4\textwidth]{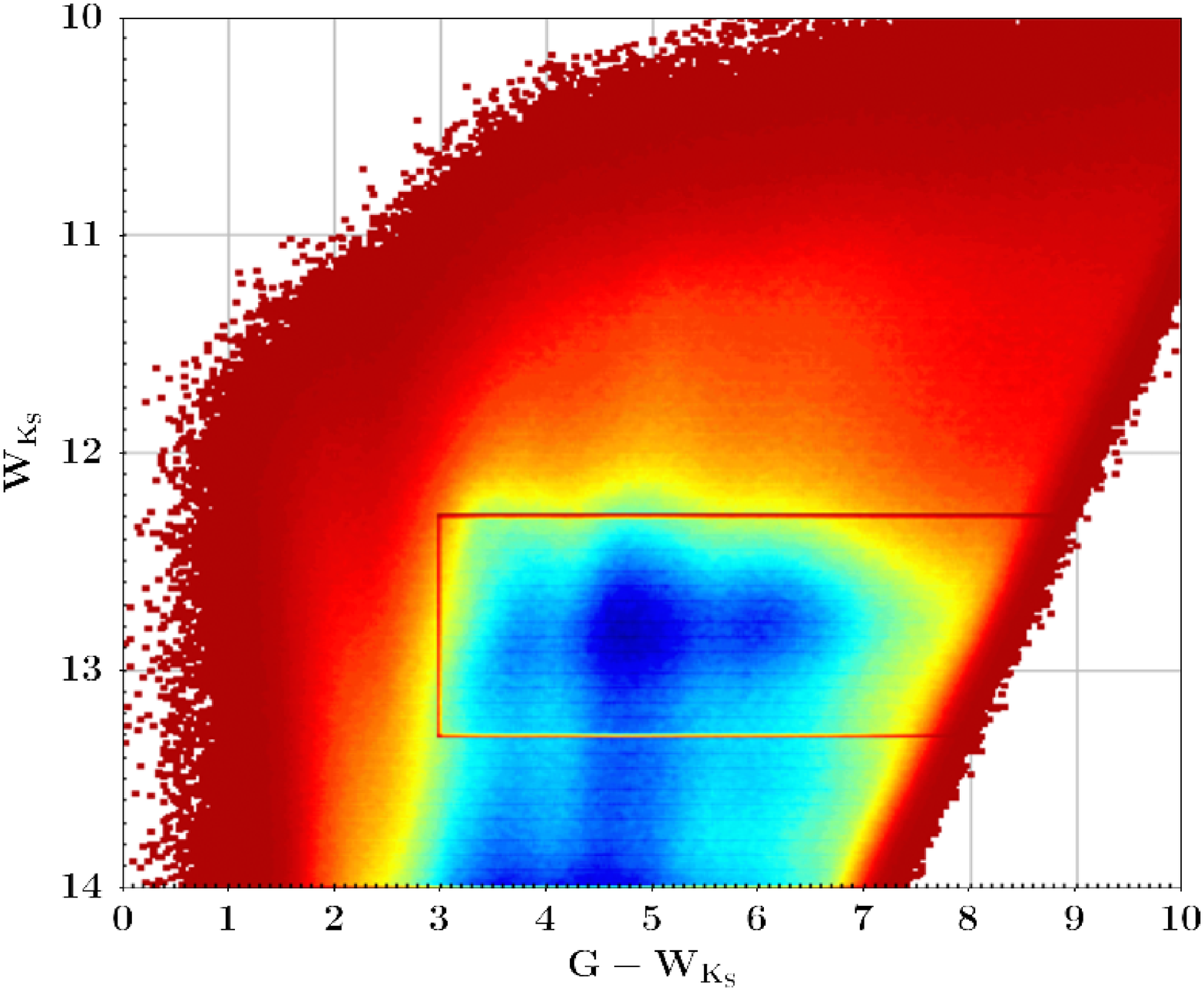}}%
\subfigure[]{%
\label{fig:wg-wk}%
\includegraphics[width=0.4\textwidth]{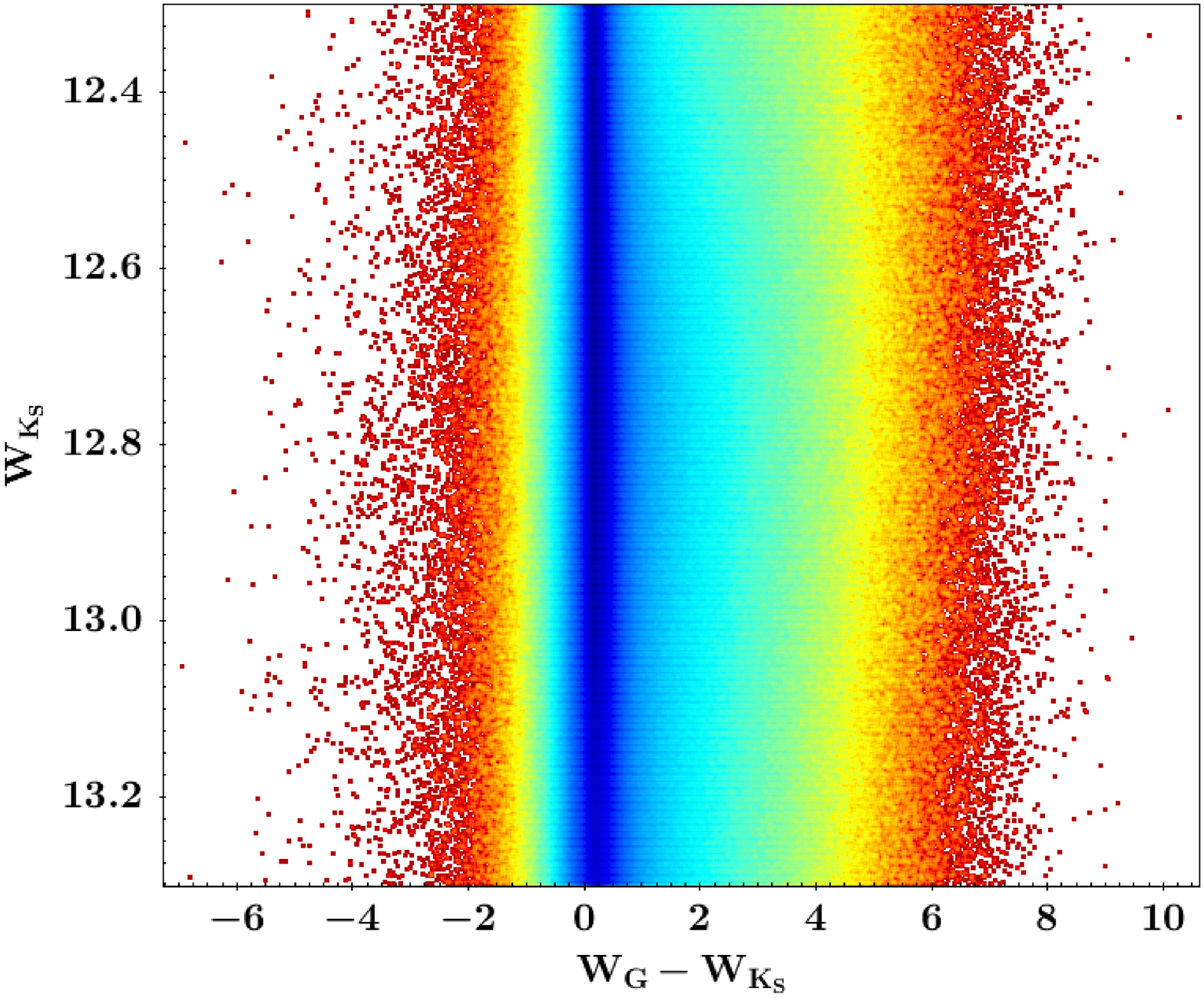}}%
\qquad
\subfigure[]{%
\label{fig:cmd g-k 2}%
\includegraphics[width=0.4\textwidth]{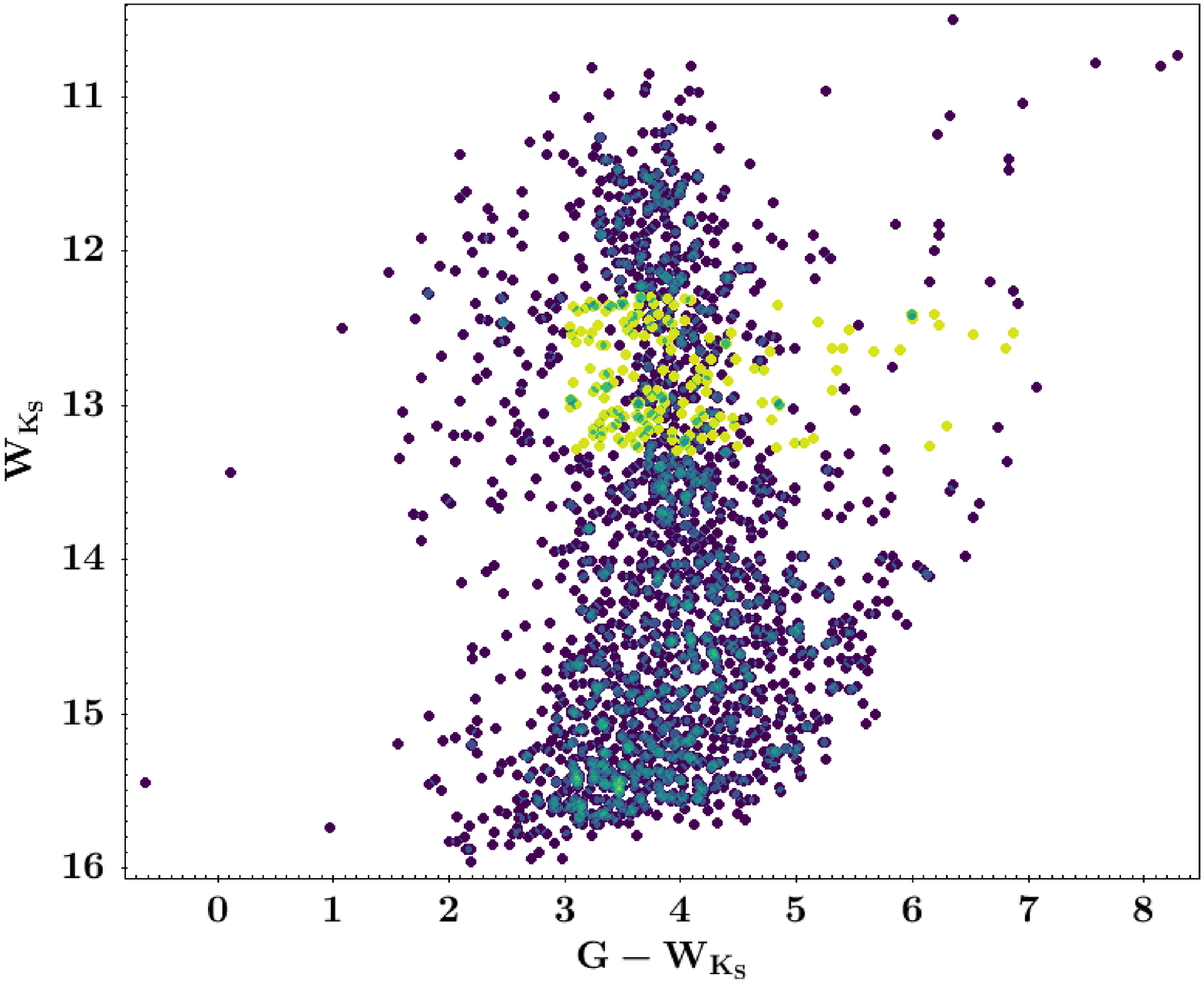}}
\subfigure[]{%
\label{fig:wg-wk 2}%
\includegraphics[width=0.4\textwidth]{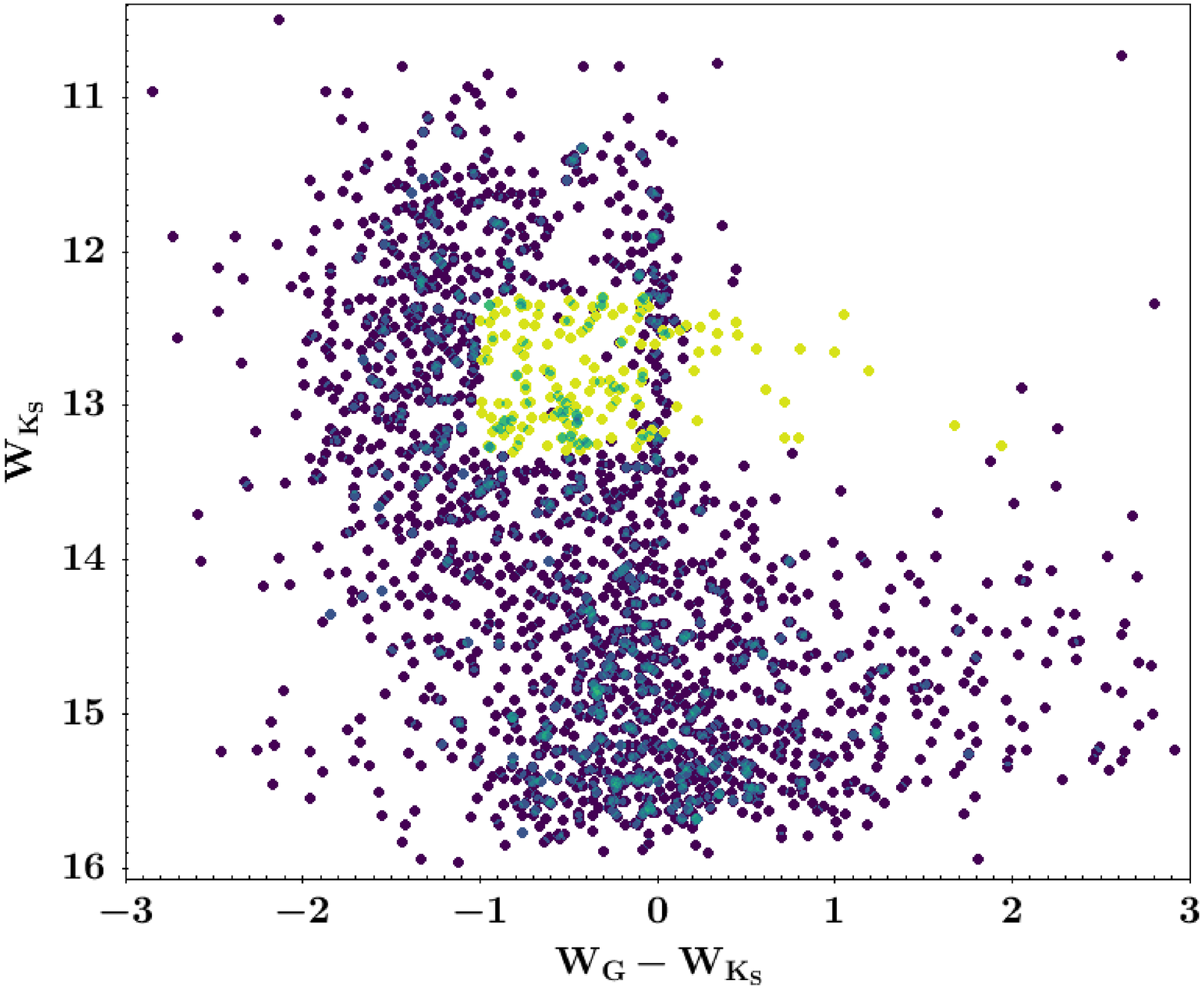}}
\caption{Near-IR CMD from VVV and Gaia using the reddening independent Wesenheit magnitudes $W_{K_S}$ and $W_G$. a) The selection box used to single out bulge RC stars is shown. b) Selected RC stars; from these, we select the ones in the range $-1<(W_G-W_{K_S})<2$. c) Same as subfigure a), but with the new sample from the VVV, the green dots correspond to the color cut shown in an enclosed box in a). d) Using the Wesenheit magnitudes, we select the ones in the range $-1<(W_G-W_{K_S})<2$.}
\end{figure}

We use the Wesenheit color-magnitude diagram (CMD) in order to select RC stars. We apply color and magnitude cuts to the VVV CMD following \citet{Minniti17} and similar color and magnitude cuts to the Gaia CMD. Combining Gaia $G$ magnitude and VVV $K_S$ magnitude, we construct a CMD and select the sources in the locus of the RC, shown enclosed in a box in Figure \ref{fig:cmd g-k}. From such sources, we select the ones in the range $-1<(W_G-W_{K_S})<2$ in the CMD shown in Figure \ref{fig:wg-wk}.

Now in order to select candidates with high proper motions in the inner Galaxy, from the VVV survey PSF catalog, we took the sources appearing in both studied epochs (2010 and 2015) and with observations in both $J$ and $K_S$ filters. The mean magnitude in $K_S$ must be $K_S<16$, in order to keep heavily reddened candidates, and maintaining a constant magnitude between the two epochs in $J$ and as well in $K_S$ (mean magnitude-magnitude in a given epoch $< 3 \sigma$).
Differently from \citet{alonso18}, now we also required that the sources had a spatial separation between both observation epochs of $0.34\arcsec <tolerance<2.5\arcsec$, since the HVS candidates should show a larger separation between the observations. 

We then match the stellar position of the Gaia DR2 and the second epoch of observation (2015) from the VVV sources with a tolerance of 0.5 arcsec, resulting in a total of $N=2,752$ sources that, as before, have accurate positions in the Galactic bulge region ($-10\degree<l<10\degree$, $-10\degree<b<5\degree$), optical and near-IR photometry and PMs. Figures \ref{fig:cmd g-k 2} and \ref{fig:wg-wk 2} show the CMD for such sample, where the light green points represent the RC stars, obtained from the same color cuts as explained before, shown in Figures \ref{fig:cmd g-k} and \ref{fig:wg-wk}. An exclusion for Gaia DR2 sources with well defined parallaxes must be done, we excluded those sources in which parallax error is less than 20\%, placing them within 500 pc. Such procedure gives us a final sample of $N_{final}=34$ RC stars.

In order to determine distances, we use the calibrations made by \citet{ruiz} accounting that the Wesenheit magnitudes are reddening-independent, so we can compute the absolute magnitudes as follows:

\begin{equation}
M_G=(0.495\pm0.009)+(1.121\pm0.128)(G-K_S-2.1)
\end{equation}

\begin{equation}
M_{K_S}=(-1.606\pm0.009)+(0.121\pm0.125)(G-K_S-2.1)
\end{equation}

And in these cases, $G=W_G$ and $K_S=W_{K_S}$. It must be noted that the calibration for Gaia photometry described above is derived from Gaia DR1 and our data is from DR2, such photometric systems are different. Nevertheless, one can derive Gaia DR1 G magnitude from DR2 G magnitude following \citep{gaia18}:
\begin{equation}
    G_{DR1}-G_{DR2}=-0.013612-0.079627(G_{BP}-G_{RP})-0.0040444(G_{BP}-G_{RP})^2+0.0018602(G_{BP}-G_{RP})^3.
\end{equation}

The distance in pc is:

\begin{equation}
d=10^{(m_i-M_i+5)/5}
\end{equation}

\noindent where $m_i=W_{K_S}; W_{G_{DR1}}$ and $M_i= M_{K_S}; M_{G_{DR1}}$.

The RC giants are good distance indicators and allow us to define a volume limited sample. All of the 34 RC stars within $-10\degree<l<10\degree$, $-10\degree<b<5\degree$, lie within $6<d<11$ kpc.

\begin{deluxetable}{ccrrrrrrrc}
\tablecaption{Gaia and VVV Photometry of Bulge RC Stars. \label{tab:table1}}
\tablehead{
\colhead{\multirow{2}{*}{ID}} &
\colhead{\multirow{2}{*}{Source ID$^a$}} & 
\colhead{${RA}$} & 
\colhead{${DEC}$} & 
\colhead{Long} & 
\colhead{Lat} & 
\colhead{$J$} & 
\colhead{$Ks$} & 
\colhead{$G_{DR1}$}
\\
 & &  
\colhead{(\degree)} & 
\colhead{(\degree)} &
\colhead{(\degree)} & 
\colhead{(\degree)} & 
\colhead{(mag)} & 
\colhead{(mag)} &
\colhead{(mag)}
}
\startdata
1	&	4043657695838288768	&	268.270	&	-31.751	&	-1.588	&	-2.827	&	14.195	&	13.261	&	17.252	\\
2	&	4050119216276193152	&	272.291	&	-29.664	&	1.938	&	-4.809	&	13.949	&	13.138	&	16.265	\\
3	&	4050891554627761536	&	273.355	&	-28.059	&	3.799	&	-4.865	&	13.670	&	12.873	&	15.649	\\
4	&	4053940195338701056	&	265.668	&	-32.979	&	-3.776	&	-1.578	&	15.156	&	13.758	&	19.423	\\
5	&	4055694466139424896	&	268.301	&	-31.314	&	-1.197	&	-2.629	&	14.253	&	13.015	&	17.639	\\
6	&	4055974493662728064	&	267.280	&	-30.608	&	-1.037	&	-1.515	&	14.216	&	13.059	&	19.056	\\
7	&	4056079565914165760	&	268.455	&	-30.941	&	-0.809	&	-2.554	&	14.583	&	13.622	&	17.471	\\
8	&	4056243191160985728	&	269.922	&	-29.962	&	0.673	&	-3.159	&	13.587	&	13.242	&	16.723	\\
9	&	4056355405826018688	&	267.689	&	-30.394	&	-0.673	&	-1.708	&	14.660	&	13.266	&	18.528	\\
10	&	4056475218139133568	&	267.952	&	-29.386	&	0.311	&	-1.389	&	14.969	&	13.284	&	19.394	\\
11	&	4056562732513987200	&	268.390	&	-29.274	&	0.602	&	-1.662	&	14.364	&	13.182	&	17.416	\\
12	&	4056575308188029056	&	268.221	&	-29.192	&	0.598	&	-1.492	&	14.386	&	13.090	&	17.958	\\
13	&	4056799093014910848	&	266.692	&	-30.180	&	-0.933	&	-0.860	&	15.229	&	13.534	&	18.118	\\
14	&	4059583335633235456	&	262.454	&	-29.028	&	-1.929	&	2.870	&	14.657	&	13.518	&	17.850	\\
15	&	4060422022558245248	&	264.260	&	-28.505	&	-0.629	&	1.824	&	14.508	&	13.059	&	18.390	\\
16	&	4060809836556331648	&	266.370	&	-27.133	&	1.523	&	0.966	&	14.286	&	13.449	&	17.272	\\
17	&	4060841348825794432	&	265.091	&	-27.820	&	0.342	&	1.570	&	14.440	&	13.056	&	18.432	\\
18	&	4060858970977014784	&	264.922	&	-27.641	&	0.414	&	1.792	&	14.635	&	13.101	&	18.411	\\
19	&	4060875566737607936	&	265.435	&	-27.529	&	0.751	&	1.465	&	14.635	&	13.150	&	18.714	\\
20	&	4061174432075039488	&	263.213	&	-28.048	&	-0.742	&	2.847	&	14.549	&	13.226	&	18.031	\\
21	&	4061331898466070528	&	264.003	&	-26.979	&	0.537	&	2.836	&	14.766	&	13.479	&	18.206	\\
22	&	4061839842728015360	&	265.130	&	-26.232	&	1.708	&	2.382	&	14.882	&	13.755	&	18.312	\\
23	&	4062483636849326080	&	270.111	&	-28.542	&	1.990	&	-2.599	&	13.669	&	13.292	&	16.562	\\
24	&	4063159634713680384	&	271.092	&	-27.073	&	3.698	&	-2.631	&	14.063	&	13.172	&	16.617	\\
25	&	4063179906906412032	&	270.221	&	-27.371	&	3.057	&	-2.104	&	13.836	&	12.953	&	16.805	\\
26	&	4064174891897213312	&	270.357	&	-25.843	&	4.445	&	-1.454	&	14.355	&	13.167	&	19.434	\\
27	&	4064524266009265024	&	272.582	&	-26.862	&	4.527	&	-3.691	&	13.633	&	13.082	&	16.970	\\
28	&	4064649163703681792	&	273.736	&	-26.349	&	5.472	&	-4.354	&	15.235	&	13.838	&	17.394	\\
29	&	4065639720555867904	&	271.476	&	-25.574	&	5.175	&	-2.200	&	14.668	&	13.502	&	17.806	\\
30	&	4066241497024032768	&	272.943	&	-24.070	&	7.137	&	-2.639	&	14.442	&	13.195	&	17.934	\\
31	&	4068160003069338112	&	265.081	&	-25.132	&	2.619	&	3.001	&	14.750	&	13.530	&	17.685	\\
32	&	4089677381280701312	&	274.847	&	-22.935	&	8.968	&	-3.638	&	13.852	&	13.267	&	16.633	\\
33	&	4110276761609748096	&	264.153	&	-24.768	&	2.479	&	3.908	&	13.421	&	12.976	&	17.491	\\
34	&	4116316451998289664	&	264.374	&	-24.293	&	2.989	&	3.992	&	14.815	&	13.729	&	18.228	\\
\enddata
\tablenotetext{a}{Gaia DR2 ID}
\end{deluxetable}

With the distance estimation and using {\tt gala} package \citep{prince-whelan} from {\tt Astropy} \citep{astropy13}, we corrected the proper motions from Gaia DR2 ($\mu_{\alpha^*}=\mu_{\alpha} cos\delta$ and $\mu_\delta$) for the reflex motion of the Sun ($(u_{\odot},v_{\odot},w_{\odot})=(11.1\pm0.075, 245\pm 9, 7.25\pm0.37) ~km~s^{-1}$), taking the default Solar motion relative to the Galactic centre as a combination for the peculiar velocity \citep{sch10} and for the circular velocity at the Solar radius \citep{bovy15}.
Given the corrected RA and DEC proper motions, we transform to proper motions in galactic coordinates ($\mu_{l^*}=\mu_{l} cosb$ and $\mu_b$) with the same package. 

\begin{deluxetable}{ccrrrrrrc}
\tablecaption{Wessenheit magnitudes and distance estimations for the Bulge RC stars. \label{tab:dist}}
\tablehead{
\colhead{\multirow{2}{*}{ID}} &
\colhead{\multirow{2}{*}{Source ID$^a$}} & 
\colhead{$W_{K_S}$} & 
\colhead{$W_{G_{DR1}}$} & 
\colhead{$M_{G_{DR1}}$} & 
\colhead{$M_{W_{K_S}}$} & 
\colhead{$d_{K_S}$} &
\colhead{$d_{G_{DR1}}$} &  
\\
 & & 
\colhead{(mag)} & 
\colhead{(mag)} & 
\colhead{(mag)} &  
\colhead{(mag)} & 
\colhead{(pc)} & 
\colhead{(pc)} &
}
\startdata
1	&	4043657695838288768	&	12.841	&	12.485	&	-2.258	&	-1.903	&	8889.420	&	8885.327	\\
2	&	4050119216276193152	&	12.772	&	12.978	&	-1.628	&	-1.835	&	8346.398	&	8342.555	\\
3	&	4050891554627761536	&	12.514	&	12.642	&	-1.716	&	-1.845	&	7443.848	&	7440.421	\\
4	&	4053940195338701056	&	13.129	&	14.806	&	0.021	&	-1.657	&	9061.674	&	9057.502	\\
5	&	4055694466139424896	&	12.458	&	12.906	&	-1.357	&	-1.806	&	7124.595	&	7121.315	\\
6	&	4055974493662728064	&	12.538	&	12.990	&	-1.352	&	-1.805	&	7389.867	&	7386.465	\\
7	&	4056079565914165760	&	13.190	&	13.104	&	-1.955	&	-1.870	&	10283.742	&	10279.007	\\
8	&	4056243191160985728	&	13.087	&	12.775	&	-2.209	&	-1.898	&	9929.270	&	9924.698	\\
9	&	4056355405826018688	&	12.638	&	12.970	&	-1.487	&	-1.820	&	7791.625	&	7788.038	\\
10	&	4056475218139133568	&	12.525	&	12.861	&	-1.482	&	-1.819	&	7394.923	&	7391.518	\\
11	&	4056562732513987200	&	12.650	&	12.885	&	-1.596	&	-1.832	&	7877.056	&	7873.429	\\
12	&	4056575308188029056	&	12.506	&	12.540	&	-1.821	&	-1.856	&	7454.419	&	7450.987	\\
13	&	4056799093014910848	&	12.771	&	13.961	&	-0.525	&	-1.716	&	7897.280	&	7893.644	\\
14	&	4059583335633235456	&	13.006	&	12.542	&	-2.379	&	-1.916	&	9646.467	&	9642.026	\\
15	&	4060422022558245248	&	12.407	&	11.731	&	-2.617	&	-1.942	&	7409.066	&	7405.655	\\
16	&	4060809836556331648	&	13.073	&	12.368	&	-2.649	&	-1.945	&	10082.708	&	10078.066	\\
17	&	4060841348825794432	&	12.433	&	12.322	&	-1.984	&	-1.874	&	7267.601	&	7264.255	\\
18	&	4060858970977014784	&	12.411	&	12.736	&	-1.495	&	-1.821	&	7020.124	&	7016.892	\\
19	&	4060875566737607936	&	12.482	&	12.649	&	-1.672	&	-1.840	&	7317.010	&	7313.641	\\
20	&	4061174432075039488	&	12.630	&	13.192	&	-1.229	&	-1.792	&	7663.788	&	7660.260	\\
21	&	4061331898466070528	&	12.899	&	13.510	&	-1.174	&	-1.786	&	8650.874	&	8646.891	\\
22	&	4061839842728015360	&	13.248	&	12.907	&	-2.242	&	-1.901	&	10712.555	&	10707.623	\\
23	&	4062483636849326080	&	13.122	&	12.667	&	-2.369	&	-1.915	&	10171.570	&	10166.887	\\
24	&	4063159634713680384	&	12.770	&	12.580	&	-2.072	&	-1.883	&	8523.819	&	8519.895	\\
25	&	4063179906906412032	&	12.556	&	12.070	&	-2.403	&	-1.919	&	7850.388	&	7846.774	\\
26	&	4064174891897213312	&	12.632	&	11.844	&	-2.742	&	-1.955	&	8269.080	&	8265.272	\\
27	&	4064524266009265024	&	12.834	&	12.566	&	-2.160	&	-1.893	&	8818.134	&	8814.074	\\
28	&	4064649163703681792	&	13.209	&	13.925	&	-1.056	&	-1.773	&	9918.750	&	9914.184	\\
29	&	4065639720555867904	&	12.977	&	13.695	&	-1.054	&	-1.773	&	8911.397	&	8907.294	\\
30	&	4066241497024032768	&	12.634	&	13.437	&	-0.959	&	-1.763	&	7574.675	&	7571.188	\\
31	&	4068160003069338112	&	12.980	&	12.263	&	-2.663	&	-1.947	&	9670.046	&	9665.594	\\
32	&	4089677381280701312	&	13.004	&	12.745	&	-2.150	&	-1.891	&	9528.832	&	9524.445	\\
33	&	4110276761609748096	&	12.775	&	12.170	&	-2.538	&	-1.933	&	8743.221	&	8739.195	\\
34	&	4116316451998289664	&	13.240	&	12.761	&	-2.396	&	-1.918	&	10756.680	&	10751.727	\\
\enddata
\tablenotetext{a}{Gaia DR2 ID}
\end{deluxetable}

While the Galactic coordinates have negligible errors, the individual distance errors are more difficult to estimate, because they are a combination of photometric errors, intrinsic RC error, and extinction errors.
The photometric error for the typical RC magnitudes considered here is $\sigma K_S=0.02$ mag \citep{Saito12}.
The error due to the intrinsic magnitude dispersion of the RC is $K_S=0.009$ mag \citep{ruiz}.
The error from the extinction corrections depends on the slope of the adopted reddening law, and on the individual stellar reddening. Taking as a dereddened mean color $(J-K_S)_0=0.68$ \citep{gonzalez12}, our RC sample yield $-0.335<E(J-K_S)<1.015$ mag.
We estimate a typical total distance modulus error of $\sigma (m-M)_0 \approx 0.02$ mag, equivalent to $\sigma D \approx 0.38$ kpc 
at the distance of the bulge, adopted here to be $D_0=8.3$ kpc \citep{dekany15}.

We compute tangential velocities for all these stars using:
$$V_T (km~ s^{-1})=4.74 \times D(pc) \times PM(arcsec~yr^{-1})$$ 

\noindent where the distances are the ones estimated from \citet{ruiz} using $W_{K_S}$ magnitudes.

From the $V_T$ equation it is clear that the errors in the tangential velocities are a combination of the distance errors described above, and the PM errors. Adopting typical Gaia PM errors at $G=17$ mag of $\sigma PM = 0.2$ mas yr$^{-1}$, yields a mean error $\sigma V_T =21.38$ km s$^{-1}$,
using the $K_S$ distance estimation for the sample.

The bulge is a very complicated region in terms of severe crowding and high reddening, and the optical Gaia PMs in this region may suffer from unknown errors. We note that the majority of the present stars are bright enough ($G<20$ mag and $K_S<14$ mag) so they should not be significantly affected by Poisson uncertainties.


\begin{figure}[h]
\centering
\includegraphics[width=0.8\textwidth]{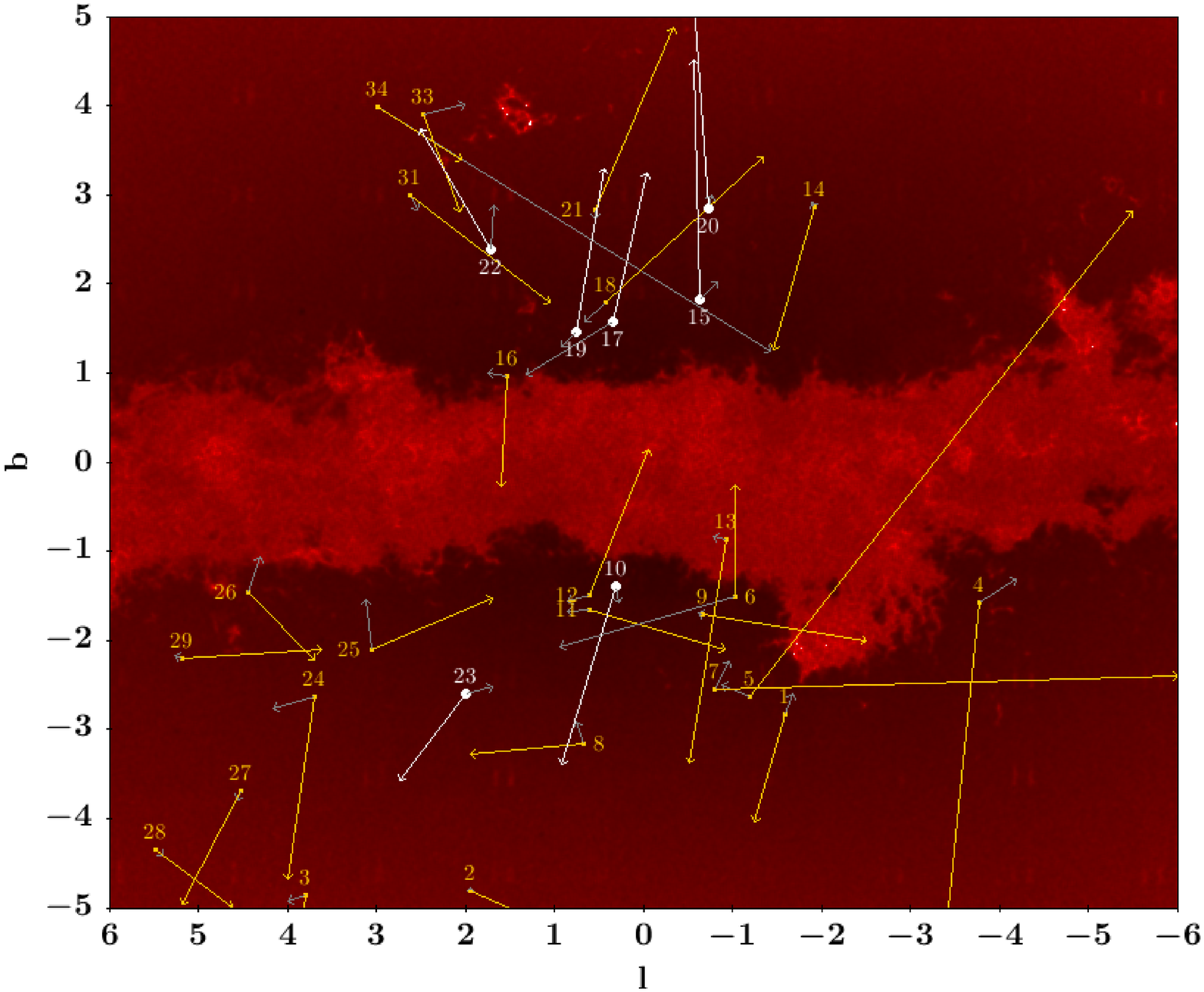}
\caption{
Density map of all RC sources with Gaia and VVV photometry. The red dots correspond to Gaia DR2 sources from the RC.
The gray vectors show the Gaia PM (scaled 5 times so they can be visible), and the yellow vectors the VVV ones; they are labelled with their ID numbers. The white vectors represent the PMs derived from the VVV for the seven HVS candidates.}
\label{fig:hvs_stream}
\end{figure}


We aim to locate HVS in the bulge ejected from the galactic centre, these would be HVS candidates ejected by binary interactions with the central SMBH \citep{hills,brown15}. Therefore we select HVS candidates that appear to radiate from the Galactic centre.
For this, we project the PM vectors, appropriately accounted for the Solar motion, to be coming from a region within 2\degree ~($\sim 300$ pc) from the Galactic centre. This box of 2\degree ~was selected in order to account for errors in the individual PMs.

Accounting for the VVV derived PMs, we found 7 stars that radiate from the Galactic centre, whose IDs are: 10, 15, 17, 19, 20, 22 and 23; their equatorial and galactic coordinates, $J$, $K_S$ and $G_{DR1}$ magnitudes are listed in Table \ref{tab:table1}; while their Wesenheit apparent and absolute magnitudes, as well as their distance estimations are listed in Table \ref{tab:dist}.
As we have the stellar positions for the first epoch of observation in 2010 and the second one in 2015, we can derive the PM from the VVV data and compare this results with those of Gaia DR2.
None of the stars that appear to radiate from the centre concur in terms of the PMs from Gaia DR2 or derived from the VVV. 

In order to verify if the usage of VVV derived PMs is a valid option, we have compared such PMs with Gaia DR2 PMs from sources with a well-behaved astrometric solution, given by the astrometric excess noise and its significance \citep{Lindegren2012}. These comparison shows us that the Gaia and VVV PMs in this regime are comparable up to 95\%, hence we can use this parameters to neglect the Gaia PM for those stars with significant astrometric excess noise and use the VVV more accurate PMs. The seven stars fulfil said condition, which make them unreliable sources in terms of the astrometric solution from Gaia DR2; assuming that the PMs derived from the VVV data are accurate, we propose these stars as HVS candidates, that are shown with white vectors in Figure \ref{fig:hvs_stream}.

\section{Discussion}
\label{sec:sec4}

The spatial distribution shows that the candidate HVS with trajectories that point directly away from the Galactic centre avoid the most reddened regions, but are concentrated to the plane (Figure \ref{fig:hvs_stream}).
Taking into account that the on-going extension of the VVV survey (VVVX)  will roughly double the areal coverage in the bulge, we can expect that the number of HVS that we will discover with the VVVX observations will also duplicate.

A qualitative comparison with models yields good agreement, because the models predict that the highest density of HVS would be in the bulge, closer to their production site. We found some excellent candidates that need to be followed-up.
A more quantitative comparison can be left for the future because of the absence of radial velocities. The \al{seven} selected candidate hypervelocity RC stars (Table \ref{tab:table2}) are prime targets for radial velocity follow-up observations. The radial velocities would allow to determine the orbits, in order to confirm if these are objects unbounded from the Milky Way.

In order to compute how long ago the HVS were ejected from the Galactic centre, we use the negative PM vectors. 
We project the PM vectors backwards, in the direction towards the Galactic centre, assuming that the modulus is correct and the direction is slightly off.


These times are listed in Table \ref{tab:table2}. Based on these times, we can estimate that there were at least seven stars ejected in the past $\sim 2.1\times 10^4$ yr, or a rate of $3.26\times 10^{-4}$ stars yr$^{-1}$.
This value is in agreement with the models of \citet{zhang}, that estimate ejection rates of $10^{-4}-10^{-5} $ stars yr$^{-1}$ and with the rate adopted by \citet{brown15} given the various theoretical results of $10^{-4}$ stars yr$^{-1}$.

\begin{deluxetable}{ccrrrrc}[h]
\tablecaption{Candidate Hypervelocity Bulge RC Stars. Tangential velocities and ejection times computed from VVV PMs. \label{tab:table2}}
\tablehead{
\colhead{\multirow{2}{*}{ID}} &
\colhead{\multirow{2}{*}{Source ID$^a$}} & 
\colhead{Distance} &
\colhead{$PM^b$} &
\colhead{$V_T$} &
\colhead{$T_{eject}$} &
\\
&
&
\colhead{(pc)} &
\colhead{($mas~yr^{-1}$)} &
\colhead{(km s$^{-1}$)} &
\colhead{(yr)}&
}
\startdata
10	&	4056475218139133568	&	7394.923	 &	208.743	$\pm 28$ &	7317.575 &	24554.129 \\
15	&	4060422022558245248	&	7409.066 &	269.609	$\pm 23$ &	9469.333 &	25766.906 \\
17	&	4060841348825794432	&	7267.601 &	171.367	$\pm 27$ &	5903.893 &	33755.978 \\
19	&	4060875566737607936	&	7317.010 &	185.743	$\pm 26$ &	6442.689&	31913.086\\
20	&	4061174432075039488	&	7663.788 &	493.208	$\pm 24$ &	17918.229&	21476.753\\
22	&	4061839842728015360	&	10712.555 &	156.447	$\pm 30$ &	7944.791 &	67448.368 \\
23	&	4062483636849326080	&	10171.570 &	123.374	$\pm 31$ &	5948.829&	95509.047\\
\enddata
\tablenotetext{a}{Gaia DR2 ID}
\tablenotetext{b}{PM errors given that the NIR positional uncertainty for VVV amounts to 0.08$\arcsec$. Further information can be found at~\url{http://apm49.ast.cam.ac.uk/surveys-projects/vista/technical/astrometric-properties}.}
\end{deluxetable}

It has been argued that there could be discrete episodes of ejections, as if, for example, a cluster of stars had a close encounter with the central SMBH \citep{Capuzzo,Fragione16}; in our sample we can identify a major ejection episode, with HVS 10, 15, 17 and 19 ejected about $3 \times 10^4$ years ago, followed by HVS 22 that has an ejection time of about $6.7 \times 10^4$ years and HVS 23 with ejection $9.5 \times 10^4$ years.

These admittedly crude results are based on small number statistics, and on assuming that the ejected stars travel at a constant velocity, so they should be taken with caution as first approximations.
The acid test for this sample would be the measurement of radial velocities, that in combination with the distances and proper motions would give us the orbital parameters for the individual stars.

\section{Conclusions}
\label{sec:sec5}

Bulge RC stars are ideal targets to search for HVS because they are numerous and their distances can be readily estimated.
We obtained a bonafide sample of 34 bulge RC giants using  near-IR data from the VVV Survey and optical data from Gaia DR2. Seven of these stars have PM vectors derived from the VVV consistent with being ejected from the vicinity of the Galactic centre, however, the PM obtained from Gaia do not concur, nevertheless, these stars have a significant astrometric excess noise within Gaia DR2 astrometric solution, making the Gaia PM unreliable, hence we propose these seven stars as HVS candidates. This would be the first detection of hypervelocity RC stars in the MW bulge.

Assuming that these candidate hypervelocity RC stars are real, we put limits to the total production rate of HVS from the Galactic centre SMBH, obtaining $N_{HVRC} \sim 3.26\times 10^{-4}$ yr$^{-1}$ (i.e. one RC star ejected every about 3000 years). Radial velocity follow-up observations are needed to confirm the hypervelocity RC star candidates, to estimate their orbital parameters, and to refine their times of ejection from the Galactic centre.

New samples of HVS are needed for constraining
the Galactic center ejection mechanism \citep{brown18}.
This work shows that it is possible to find HVS in the bulge, close to their production site. If these stars are proven to be HVS, they would be the fastest ones found to date.
It also represents the first step for mapping the distribution of bulge HVS. 
The recently started observations for the VVVX survey cover a larger area, significantly extending the map.
Interestingly, the LSST \citep{ivezic} and the WFIRST \citep{Spergel15,Stauffer} would be promising future tools to identify even larger (perhaps by orders of magnitude) samples of bulge HVS and measure more precisely their production rate.

\acknowledgments 

We gratefully acknowledge  data from the ESO Public  Survey program ID
179.B-2002  taken with  the  VISTA telescope,  and  products from  the
Cambridge Astronomical Survey Unit (CASU).  
D.M.  gratefully acknowledges support provided by the
BASAL  Center  for  Astrophysics and  Associated  Technologies  (CATA)
through grant PFB-06. D.M. and J.A.-G. also acknowledge support provided by the Ministry for the  Economy, Development
and Tourism, Programa Iniciativa  Cient\'ifica Milenio grant IC120009,
awarded to  the Millennium Institute  of Astrophysics (MAS), and
from project Fondecyt No. 1170121. A.L. acknowledges support by CONACyT M\'exico for the graduate scholarship, fellowship number 672829.
We are also very grateful for the hospitality of the Vatican Observatory, 
where this work was started.




\end{document}